\documentclass[aps,prf,showpacs,superscriptaddress,groupedaddress]{revtex4}  
\usepackage{graphicx,color}  
\usepackage{bm}        
\usepackage{amssymb}   

\begin{document}

\title{Rico and the jets: Direct numerical simulations of turbulent liquid jets}

\author{C. R. Constante-Amores}
\affiliation{Department of Chemical Engineering, Imperial College London, South Kensington Campus, London SW7 2AZ, United Kingdom}
\email{crc15@imperial.ac.uk}
\author{L.~Kahouadji}
\affiliation{Department of Chemical Engineering, Imperial College London, South Kensington Campus, London SW7 2AZ, United Kingdom}
\author{A.~Batchvarov}
\affiliation{Department of Chemical Engineering, Imperial College London, South Kensington Campus, London SW7 2AZ, United Kingdom}
\author{S.~Shin}
\affiliation{Department of Mechanical and System Design Engineering, Hongik University, Seoul 121-791, Republic of Korea}
\author{J.~Chergui} 
\affiliation{Laboratoire d'Informatique pour la M\'ecanique et les Sciences de l'Ing\'enieur (LIMSI), Centre National de la Recherche Scientifique (CNRS), Universit\'e Paris Saclay, B\^at. 507, Rue du Belv\'ed\`ere, Campus Universitaire, 91405 Orsay, France}
\author{D.~Juric}
\affiliation{Laboratoire d'Informatique pour la M\'ecanique et les Sciences de l'Ing\'enieur (LIMSI), Centre National de la Recherche Scientifique (CNRS), Universit\'e Paris Saclay, B\^at. 507, Rue du Belv\'ed\`ere, Campus Universitaire, 91405 Orsay, France}
\author{O.~K.~Matar}
\affiliation{Department of Chemical Engineering, Imperial College London, South Kensington Campus, London SW7 2AZ, United Kingdom}

\date{\today}

\begin{abstract}
This paper is associated with a poster winner of a 2019 American Physical Society’s Division of Fluid Dynamics (DFD) Milton van Dyke Award for work presented at the DFD Gallery of Fluid Motion. The original poster is available online at the Gallery of Fluid Motion, https://doi.org/10.1103/APS.DFD.2019.GFM.P0020
\end{abstract}

\maketitle

\begin{figure}
\includegraphics[trim=1 1 1 420, clip=true,width=\linewidth]{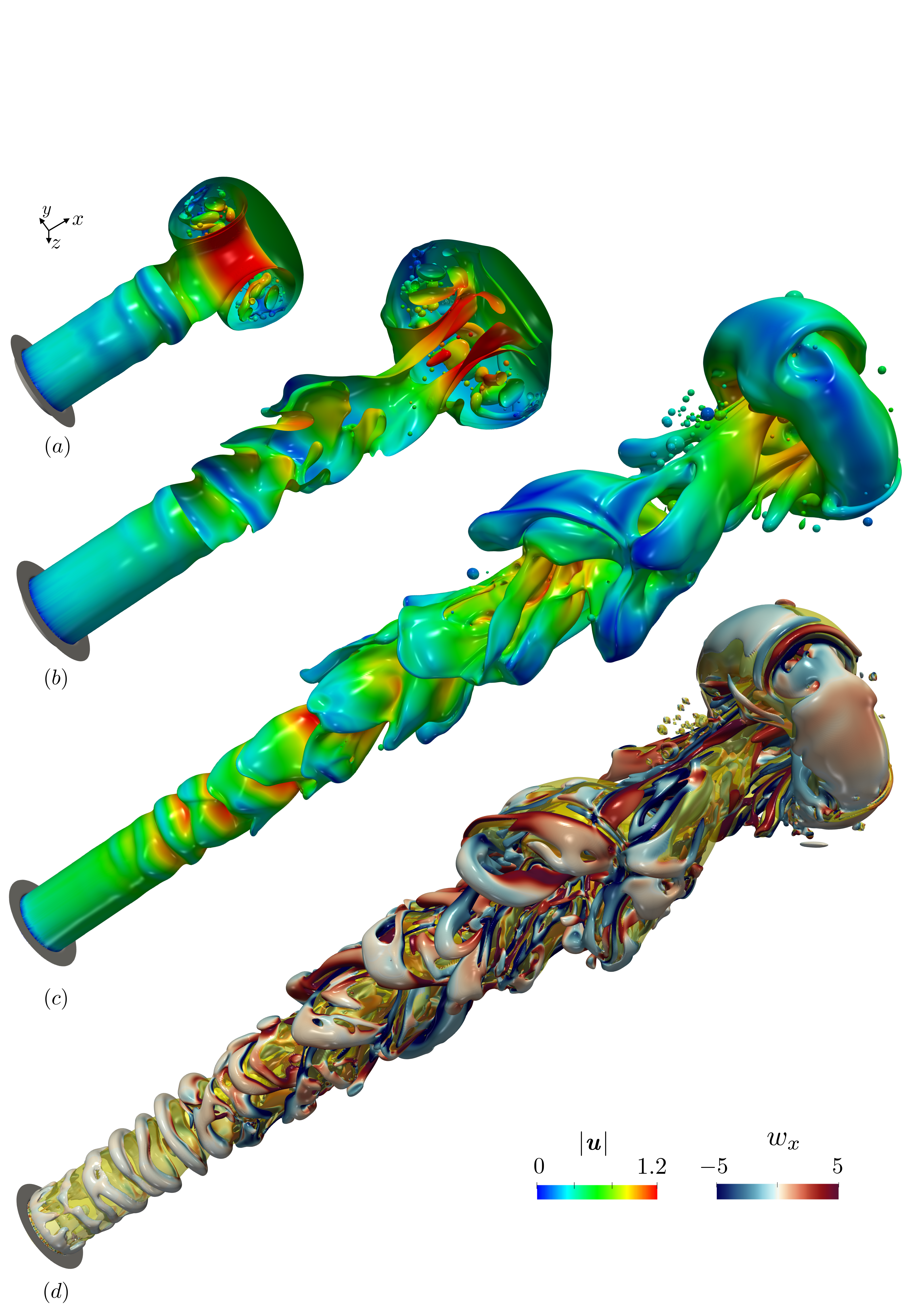} 
\caption{ \label{figure} Spatio-temporal evolution of the interface in the injection of a water jet into a stagnant viscous silicone-oil 
at $t= (7.25, 12.05, 28.97) $ corresponding to (a), (b) and (c), respectively, when $Re=6530$. (d) Illustration of the coherent vortical structures through the Q-criterion  close to the free-surface  (coloured in yellow) at $t=28.97$. The vortical structures have been coloured by the value of the vorticity in the streamwise direction. In the vorticity representation, blue and red colour represent vortical structures with counter-clockwise and clockwise rotation, respectively. All variables are dimensionless quantities.}
\end{figure}

The breakup of an interface into a cascade of droplets and their subsequent coalescence is a generic problem of central importance to a large number of industrial settings. Examples of these applications include the atomisation of propellants in engines, the formation of droplets in injectors, mixers, separators, and the generation of droplets in multiphase flow regime transitions \citep{Lin_arfm_1998, Eggers_rpp_2008, Eggers_rmp_1997}. In all of these situations, it is important to predict the evolving droplet size distribution that results from a competition between breakup and coalescence, which are influenced by a range of multi-scale physics; this includes the interaction of turbulence with interfaces, capillarity, viscosity, and gravity. Therefore, it is unsurprising that the breakup of liquid jets during injection (i.e. atomisation) has received great scientific interest \citep{ibarra_jfm_2020,Desjardins_as_2010,Marmottant_jfm_2004,Jarrahbashi_jfm_2016,Lasheras_arfm_2000}.

To the best of our knowledge, the transient dynamics of turbulent liquid/liquid systems have not been reported in the literature. Temporal instabilities and the resulting spatio-temporal interfacial structures are predicted by solving the full three-dimensional two-phase Navier-Stokes system in the context of a  hybrid front-tracking/level-set method \citep{Shin_jcp_2002, Shin_jmst_2017,Shin_jcp_2018}. We consider a cylindrical nozzle with diameter $D=4$ mm injecting water with density $\rho_{_w}$ and dynamic viscosity $\mu_{_w}$. This water jet enters progressively into the computational domain of size $20D \times 4D \times 4D$, initially filled with a stagnant viscous silicone-oil of density $\rho_{_{so}}$ and viscosity $\mu_{_{so}}$. The surface tension is taken to be that of oil and water (e.g.  $\sigma=35.1$ mN/s) \citep{Ibarra_2017}. The Reynolds number is defined as $Re=\rho_{_{w}} U_{jet} D / \mu_{_{w}}$ and fixed to the value of $Re= 6530$. The domain has been divided into $48 \times 6 \times 6$ subdomains where each subdomain contains a Cartesian structured grid of $64^3$ cells, accounting for a global structured mesh grid of $3072 \times 384 \times 384$. This mesh is sufficiently large to resolve the relevant turbulent length-scales and interfacial singularities (e.g. pinch-off and coalescence).

We use a new solver for massively parallel simulations of fully three-dimensional multiphase flows \cite{Shin_jmst_2017}, able to run on a variety of computer architectures, wholly written in Fortran 2008 and adopting an algebraic domain decomposition strategy for parallelization with MPI. The fluid interface solver is based on a parallel implementation of the Level Contour Reconstruction Method (LCRM) which is an adaptation of our high fidelity hybrid front tracking/level set method, able to handle highly deforming interfaces with complex topology changes \citep{ligament,bursting_bubble,falling_films}. This code uses parallel GMRES and multigrid iterative solvers suited to solve the linear systems arising from the implicit solution of the fluid velocities and pressure. More details on the numerical techniques can be found in \citet{Shin_jmst_2017,Shin_jcp_2018}.

The spatio-temporal evolution of the interfacial dynamics is shown in Fig. \ref{figure}.  At early injection times, large capillary pressure is generated near the leading edge, due to local interfacial curvature, leading to a radially driven flow.
This capillary-induced flow together with the viscous resistance from the stagnant phase yields the formation of a leading-top mushroom-like structure (see Fig. \ref{figure}a). This structure covers an internal interfacial toroidal sheet whose thickness reduces over time to generate the formation of holes, which expand radially to form ligaments, and eventually entrapped droplets. As time evolves, the free surface behind the leading structure adopts the shape of a `cylinder' which undergoes a Kelvin-Helmholtz instability (KH) to give the formation of initial corrugations or capillary waves on the free-surface. The KH instabilities on the free surface are triggered by the parallel motion of fluids at different velocities and are amplified by the pulsatile injection.

The interfacial dynamics of the jet can be explained by coupling the vorticity $ \omega =\bigtriangledown  \times \textbf{u}$ with the interfacial location. During the early stages and close to the injection point, the streamwise vorticity field $\omega_{x}$ is characterised by values which are two orders of magnitude smaller than the azimuthal vorticity $\omega_{\theta}$.  As the flow evolves downstream, $\omega_{x}$ becomes comparable in magnitude with $\omega_{\theta}$, leading to the deformation of axisymmetric KH vortex rings in the streamwise direction adopting a new hairpin shape. These hairpin vortices trigger the formation of interfacial lobes which are stretched downstream (from outer hairpin vortices) and upstream (from inner hairpin vortices) to eventually obtain a hairpin shape (see Fig. \ref{figure}b,c).
We have used the Q-criterion to visualise the three-dimensional nature of the vortical structure (shown in Fig. \ref{figure}d) \citep{Hunt_CTR_1988}. The topological shape of the vortex resembles the instantaneous hairpin-like vortical structures reported in experiments and numerical simulations of \citep{zhou_jfm_1999,Zandian_jfm_2018}.  Outer hairpin vortices are observed clearly, whereas the inner hairpin vortices are localised underneath the interface. Additionally, a vortex-cap covers the leading mushroom structure.

\subsection*{Acknowledgements}

This work is supported by the Engineering \& Physical Sciences Research Council, United Kingdom, through a studentship for RCA in the Centre for Doctoral Training on Theory and Simulation of Materials at Imperial College London funded by the EPSRC (EP/L015579/1) (Award reference:1808927), and through the EPSRC MEMPHIS (EP/K003976/1) and PREMIERE (EP/T000414/1) Programme Grants. OKM also acknowledges funding from PETRONAS and the Royal Academy of Engineering for a Research Chair in Multiphase Fluid Dynamics. We also acknowledge HPC facilities provided by the Research Computing Service (RCS) of Imperial College London for the computing time. DJ and JC acknowledge support through computing time at the Institut du Developpement et des Ressources en Informatique Scientifique (IDRIS) of the Centre National de la Recherche Scientifique (CNRS), coordinated by GENCI
(Grand Equipement National de Calcul Intensif) Grant 2020 A0082B06721. The numerical simulations were performed with code BLUE (\citet{Shin_jmst_2017}) and the visualisations have been generated using ParaView.


\end{document}